# Superconductivity at 40 K in lithiation-processed [(Fe,Al)(OH)$_2$][FeSe]$_{1.2}$ with a layered structure

Guobing Hu,[1,2] Mengzhu Shi,[1] Wenxiang Wang,[1] Changsheng Zhu,[1] Zeliang Sun,[1] Jianhua Cui,[1] Weizhuang Zhuo,[1] Fanghang Yu,[1] Xigang Luo,[1] and Xianhui Chen[1,3,4,5]*

[1]Key Laboratory of Strongly-coupled Quantum Matter Physics, Chinese Academy of Sciences, and Hefei National Laboratory for Physical Sciences at Microscale, and Department of Physics, University of Science and Technology of China, Hefei, Anhui 230026, China.

[2]Physical Science and Technology College, Yichun University, Yichun 336000, China

[3]Collaborative Innovation Center of Advanced Microstructures, Nanjing University, Nanjing 210093, China

[4]CAS Center for Excellence in Superconducting Electronics (CENSE), Shanghai 200050, China

[5]CAS Center for Excellence in Quantum Information and Quantum Physics, Hefei, Anhui 230026, China

**ABSTRACT:** Exploration of new superconductors has always been one of the research directions in condensed matter physics. We report here a new layered heterostructure of [(Fe,Al)(OH)$_2$][FeSe]$_{1.2}$, which is synthesized by the hydrothermal ion-exchange technique. The structure is suggested by a combination of X-ray powder diffraction and the electron diffraction (ED). [(Fe,Al)(OH)$_2$][FeSe]$_{1.2}$ is composed of the alternating stacking of tetragonal FeSe layer and hexagonal (Fe,Al)(OH)$_2$ layer. In [(Fe,Al)(OH)$_2$][FeSe]$_{1.2}$, there exists mismatch between the FeSe sub-layer and (Fe,Al)(OH)$_2$ sub-layer, and the lattice of the layered heterostructure is quasi-commensurate. The as-synthesized [(Fe,Al)(OH)$_2$][FeSe]$_{1.2}$ is non-superconducting due to the Fe vacancies in the FeSe layer. The superconductivity with a $T_c$ of 40 K can be achieved after a lithiation process, which is due to the elimination of the Fe vacancies in the FeSe layer. The $T_c$ is nearly the same as that of (Li,Fe)OHFeSe although the structure of [(Fe,Al)(OH)$_2$][FeSe]$_{1.2}$ is quite different from that of (Li,Fe)OHFeSe. The new layered heterostructure of [(Fe,Al)(OH)$_2$][FeSe]$_{1.2}$ contains an iron selenium tetragonal lattice interleaved with a hexagonal metal hydroxide lattice. These results indicate that the superconductivity is very robust for FeSe-based superconductors. It opens a path for exploring superconductivity in iron-base superconductors.

## INTRODUCTION

Exploration of new superconductors has always been one of the research directions for physicists since the discovery of superconductivity in 1911. Iron-based superconductors have attracted much attention over the past years due to its abundant structural and physical properties.[1-5] The tetragonal FeSe adopts the layered anti-PbO structure, and the $T_c$ of FeSe can be increased from 8.5 K to above 40 K by intercalation of alkali metal ions or small molecules into the adjacent FeSe layers.[6-8] Recently, an air-stable FeSe-derived superconductor Li$_{0.8}$Fe$_{0.2}$OHFeSe, which contains the alternating tetragonal FeSe layer and tetragonal (Li,Fe)OH layer, is discovered with $T_c$ ~ 43 K.[9-11] Later research indicates the Fe vacancies in the FeSe layer can be filled by lithiation process. Moreover, the Fe vacancies concentrations in the FeSe layer can influence the $T_c$ of Li$_{1-x}$Fe$_x$(OH)Fe$_{1-y}$Se.[12,13] The mechanism of iron-based high-$T_c$ superconductivity is still unclear, and discovery of new iron-based superconductor to investigate the underlying mechanism of iron-based superconductors is meaningful.

Misfit layer structure compounds, generally described as (MX)$_n$[TX$_2$]$_m$, where M =Sn, Pb, Sb, Bi, or a lanthanide, T=Ti, V, Nb, Ta, or Cr, X=S or Se, n=1.1-1.2, m=1,2, have an unusual structure which are formed by pseudo-hexagonal TX$_2$ sandwiches and pseudo-tetragonal MX layer[14,15]. Similarly, tochilinite, a natural mineral with the formula [(Mg,Fe,Al)(OH)$_2$]$_n$[Fe$_{1-x}$S]$_m$, represents a mineral group which contains the similar structure. Tochilinite is composed of the alternating stacking of mackinawite-like Fe$_{1-x}$S layer and metal hydroxide layer with brucite-like structure.[16-18] In tochilinite, the lattices of mackinawite-like tetragonal Fe$_{1-x}$S and hexagonal metal hydroxide is quasi-commensurate. [16-18] The mackinawite-like Fe$_{1-x}$S layer is formed by the edge-shared FeS$_4$ tetrahedra, which is considered to carry negative charges due to the Fe vacancies in the Fe$_{1-x}$S layer. The metal hydroxide layer is formed by the edge-shared M(OH)$_6$ octahedra, which is considered to carry positive charges.[19,20] The structure of the mackinawite-like Fe$_{1-x}$S is the same as that of β-FeSe. Inspired by natural tochilinite which is formed by a tetragonal lattice interleaved with a hexagonal lattice and (Li,Fe)OHFeSe which is formed by a tetragonal lattice interleaved with a tetragonal lattice, we try to synthesize new iron-based superconductors which maybe composed by the alternating stacking of tetragonal iron selenium layer and metal hydroxide layer with hexagonal brucite-like structure.

Here, we report a new layered material, $[(Fe,Al)(OH)_2][FeSe]_{1.2}$ synthesized through a hydrothermal ion-exchange technique. $[(Fe,Al)(OH)_2][FeSe]_{1.2}$ is formed by alternating stacking of tetragonal anti-PbO-type FeSe layer and hexagonal $(Fe,Al)(OH)_2$ layer. The electron diffraction (ED) pattern of the $[(Fe,Al)(OH)_2][FeSe]_{1.2}$ is very similar to the ED pattern of the natural tochilinite. It implies that their structures are very similar to each other. The as-synthesized $[(Fe,Al)(OH)_2][FeSe]_{1.2}$ is non-superconducting due to the Fe vacancies in the FeSe layer. Powder XRD refinement results show that large amount of Fe vacancies exist in the FeSe layer.

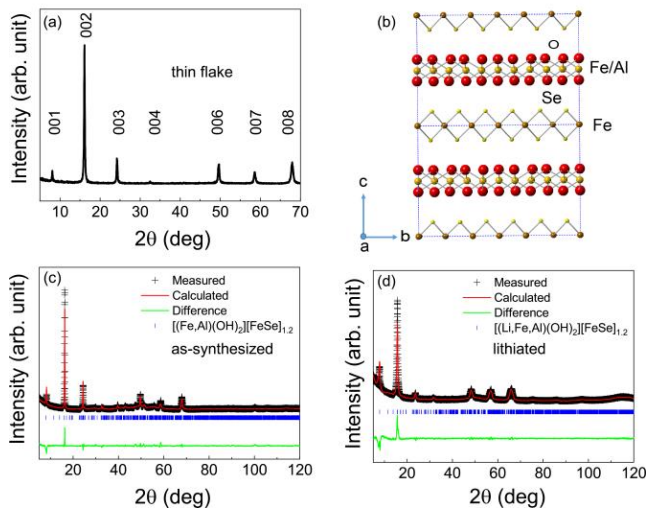

**Figure 1**. (a) XRD pattern taken from the as-synthesized $[(Fe,Al)(OH)_2][FeSe]_{1.2}$ polycrystal thin flake sample. (b) The schematic view of the crystal structure of $[(Fe,Al)(OH)_2][FeSe]_{1.2}$. (c)(d) Measured (crosses) and calculated (red solid line) powder XRD pattern for the as-synthesized $[(Fe,Al)(OH)_2][FeSe]_{1.2}$, and lithiated $[(Fe,Al)(OH)_2][FeSe]_{1.2}$, respectively. Bragg peak positions are indicated by short vertical bars below the XRD patterns. The bottom of the Figure shows the differences between measured and calculated intensities.

These vacancies can be eliminated by lithiation process using n-butyl lithium (n-BuLi) as the lithiation reagent. The magnetic susceptibility and transport measurements confirm the bulk superconductivity with transition temperature ($T_c$) of 40 K after the large amount of Fe vacancies in the FeSe layer are eliminated. The $T_c$ of $[(Fe,Al)(OH)_2][FeSe]_{1.2}$ is nearly the same as that of (Li,Fe)OHFeSe even though the structure of $[(Fe,Al)(OH)_2][FeSe]_{1.2}$ is quite different from that of (Li,Fe)OHFeSe. Our finding indicates that the layered heterostructure of $[(Fe,Al)(OH)_2][FeSe]_{1.2}$ contains a tetragonal lattice interleaved with a hexagonal lattice. It opens a path for exploring superconductivity in other related systems.

## EXPERIMENTAL SECTION

**Material Synthesis.** The $[(Fe,Al)(OH)_2][FeSe]_{1.2}$ is synthesized by a hydrothermal ion-exchange technique using an insulating $K_{0.8}Fe_{1.6}Se_2$ single crystal as a matrix. Firstly, 0.003-0.005 mol of selenourea (Alfa Aesar, 99.97% purity), 0.0035 mol of nano iron powder (macklin, 99.9% purity, 50 nm), 0.0015 mol of $Al(OH)_3$ (macklin, 99.99% purity), 0.008-0.015 mol of KOH (macklin, 99.99% purity) and several pieces of $K_{0.8}Fe_{1.6}Se_2$ single crystals were mixed with 5 ml de-ionized water and loaded into a stainless steel autoclaves of 25 ml capacity with Teflon liners. The $K_{0.8}Fe_{1.6}Se_2$ single crystal was synthesized by elements of Fe, Se and K, which had been reported elsewhere.[11] The autoclave was tightly sealed and heated at 110 -140 °C for 96 hours. The ion-exchange synthesized $[(Fe,Al)(OH)_2][FeSe]_{1.2}$ thin flakes were washed by de-ionized water and ethyl alcohol for several times, then the thin flakes were dried under vacuum and stored in an argon-filled glovebox. The atomic ratio of Fe : Se : Al is determined to be 1.42 : 1 : 0.22 by inductively coupled plasma-atomic emission spectroscopy (ICP-AES), with an instrument error of 10 %. The as-synthesized samples were subsequently subjected to lithiation process in which the thin flakes were immersed in a solution of hexane using n-butyl lithium (n-BuLi) as the lithiation reagent. The lithiation reactions were carried out for 7 hours in an argon-filled glovebox with the dilute n-BuLi (0.22 mol/L) at room temperature. (**CAUTION:** the excess unreacted n-BuLi is very dangerous and must be dealed with carefully.)

**Material Characterization.** The structure of $[(Fe,Al)(OH)_2][FeSe]_{1.2}$ is characterized by a powder X-ray diffractometer (SmartLab-9, Rigaku Corp.) with Cu Kα radiation and a fixed graphite monochromator. The lattice parameters and the crystal structure are refined using the Rietveld method with the program GSAS package. The sample is easy to decompose during the grinding process at room temperature. In order to get polycrystalline sample to refine the crystal structure, the sample is ground in liquid nitrogen in an argon-filled glovebox. (**In particular**, liquid nitrogen is very dangerous, antifreeze measures must be done during the grinding process to avoid frostbite.) The resistivity measurement is carried out by Physical Properties Measurement System (PPMS) with standard four-terminal method. The magnetic susceptibility is measured by a SQUID magnetometer (Quantum Design MPMS-5). Electron diffraction patterns are obtained using a JEOL 2010 TEM at an acceleration voltage of 200 KV.

## RESULTS AND DISCUSSIONS

The as-derived $[(Fe,Al)(OH)_2][FeSe]_{1.2}$ polycrystal thin flake roughly inherits the original shape of its precursor $K_{0.8}Fe_{1.6}Se_2$ single crystal. Figure 1a shows the room temperature x-ray diffraction (XRD) pattern of the thin flake. It indicates that the polycrystal thin flake displays (00l)-preferential orientation. The first peak corresponds to a $d$-spacing of ∼ 11.0 Å, which is significantly greater than c-axis lattice parameters of (Li,Fe)OHFeSe. The lattice parameters of (Li,Fe)OHFeSe is about 9.3 Å for c-axis.[10] Considering that the $d$-spacing of natural tochilinite is ∼ 10.72 Å along c direction,[16,18] we propose a possible crystal structure of the thin flake as shown in Figure 1b, combined with the result of ICP-AES. The schematic diagram shows that the thin flake maybe composed by the alternating stacking of tetragonal anti-PbO-type FeSe layer and hexag-onal (Fe,Al)OH$_2$ layer, which is similar to the structure of natural tochilinite.[16,18] In order to determine the crystal structure of the polycrystal thin flake, the sample is ground slowly in liquid nitrogen in an argon-filled glovebox. Structural analysis was carried out on the as-synthesized [(Fe,Al)(OH)$_2$][FeSe]$_{1.2}$ polycrystals by Rietveld refinement with the programs GSAS package[21,22], showing that the as-synthesized sample has a triclinic (pseudomonoclinic) structure with space group of $C$ 1 like the structure of tochilinite,[16,18] as shown in Figure 1c. Details of the data collection are provided in Table 1, and crystallographic parameters from the powder XRD refinement of the as-synthesized (Fe,Al)(OH)$_2$FeSe are listed in Table S1 in supporting information. The room-temperature lattice parameters are 5.5414(3) Å for $a$-axis, 15.393(7) Å for $b$-axis, 11.0879(4) Å for $c$-axis, 93.727(9)° for $\beta$ and 90° for $\alpha$ and $\gamma$. The lattice parameters of the natural tochilinite are 5.2-5.5 Å for $a$-axis, 15.3-15.9 Å for $b$-axis, 10.7-10.9 Å for $c$-axis, 93.6-95.8° for $\beta$ and 90° for $\alpha$ and $\gamma$.[16,23] Compared with the lattice parameters of the natural tochilinite, the $a$-axis, $b$-axis and $\alpha$, $\beta$, $\gamma$ are consistent, while the $c$-axis is enlarged from 10.72 Å[16,18] to 11.088(0) Å. The reason of the $c$-axis expansion may ascribe to the $c$-axis lattice parameters of FeSe crystal is larger than that of mackinawite-like FeS crystal. For the $c$-axis lattice parameters, the FeSe is 5.53 Å, and the mackinawite-like FeS is 5.03 Å, respectively. The $a$-axis lattice parameters of FeSe and mackinawite-like FeS are 3.77 Å and 3.68 Å, respectively, which may be the reason for the $a$-axis and $b$-axis lattice parameters of the as-synthesized [(Fe,Al)(OH)$_2$][FeSe]$_{1.2}$ sample are consistent with the natural tochilinite. The molecular formula of the as-synthesized [(Fe,Al)(OH)$_2$][FeSe]$_{1.2}$ is determined to be [Fe$_{0.77}$Al$_{0.23}$(OH)$_2$][Fe$_{0.78}$Se]$_{1.2}$ by the refinement, which is in accordance with the ICP-AES result. The ratio of Fe : Al in the hydroxide layer and the amount of Fe vacancies in the FeSe layer are all within a reasonable range compared with the natural tochilinite.[18,23] In addition, there exists ~ 22% Fe vacancies in the FeSe layer. This is such a large amount of Fe vacancies that the superconductivity should be killed in the as-synthesized sample. In Li$_{1-x}$Fe$_x$(OH)Fe$_{1-y}$Se, the lithiation process can effectively eliminate the Fe vacancies in the FeSe layer, consequently inducing superconductivity.[12,13] So we try to eliminate the Fe vacancies in the FeSe layer of the as-synthesized [(Fe,Al)(OH)$_2$][FeSe]$_{1.2}$ using n-butyl lithium (n-BuLi) in hexane. The XRD pattern of the lithiated sample is displayed in Figure 1d. Details of the data collection are provided in Table 1, and crystallographic parameters from the powder XRD refinement of the lithiated [(Fe,Al)(OH)$_2$][FeSe]$_{1.2}$ are listed in Table S2 in supporting information. The room-temperature lattice parameters are 5.5300(7) Å for $a$-axis, 15.569(2) Å for $b$-axis, 11.3652(9) Å for $c$-axis, 94.582(13)° for $\beta$ and 90° for $\alpha$ and $\gamma$. The $c$-axis of the lithiated sample is expanded compared with the as-synthesized sample, which is in accordance with the lithiation process in Li$_{1-x}$Fe$_x$(OH)Fe$_{1-y}$Se.[12,13] The molecular formula of the lithiated sample is determined to be [Li$_{0.15}$Fe$_{0.62}$Al$_{0.23}$(OH)$_2$][Fe$_{0.97}$Se]$_{1.2}$ by the Rietveld refinement. The mechanism how Fe vacancies are filled during lithiation process should be the same as the lithiation process in Li$_{1-x}$Fe$_x$(OH)Fe$_{1-y}$Se.[12] Li ions occupy the site of Fe in the (Fe,Al)(OH)$_2$ layer, and the expelled Fe ions by Li migrate to the FeSe layer to fill in the Fe vacancies.[12] The molecular formulas of the as-synthesized sample and the lithiated sample are consistent with the above explanation within the range of experimental error. Thus, the lithiated sample displays the decrease of Fe content in the (Fe,Al)(OH)$_2$ layer and the increase of Fe content in the FeSe layer with the ratio of Fe : Se to be 1 : 1, just like the case of the lithiated (Li,Fe)OHFeSe.[12,24] In addition, the ion radii of Li$^+$ is about 0.59 Å[25], the change of c-axis lattice parameters before and after lithiation is less than 0.3 Å, so intercalation of Li$^+$ between two layers may not occur just like the situation of the lithiation of (Li,Fe)OHFeSe.[12,13] More discussion about the possibility of Li$^+$ intercalation between the (Fe,Al)OH$_2$ layer and FeSe layer is supplied in the supplementary materials. Considering that there are ∼ 20% Fe vacancies in the insulating K$_{0.8}$Fe$_{1.6}$Se$_2$ single crystal,[26] we also try to synthesize [(Fe,Al)(OH)$_2$][FeSe]$_{1.2}$ polycrystals by hydrothermal method, without using the insulating K$_{0.8}$Fe$_{1.6}$Se$_2$ single crystal as a template. Unfortunately, we have not found superconductivity in the [(Fe,Al)(OH)$_2$][FeSe]$_{1.2}$ polycrystals directly synthesized by the hydrothermal method. No superconductivity is observed in the [(Fe,Al)(OH)$_2$][FeSe]$_{1.2}$ polycrystals due to the Fe vacancies in the FeSe layer, and the detail results are shown in Figure S1 and Figure S2 in supporting information.

**Table 1. Data collection Parameters for the as-synthesized [(Fe,Al)(OH)$_2$][FeSe]$_{1.2}$ and the lithiated [(Fe,Al)(OH)$_2$][FeSe]$_{1.2}$**

| | as-synthesized (Fe,Al)(OH)$_2$[FeSe]$_{1.2}$ | lithiated [(Fe,Al)(OH)$_2$][FeSe]$_{1.2}$ |
|---|---|---|
| empirical formula | [Fe$_{0.77}$Al$_{0.23}$(OH)$_2$][Fe$_{0.78}$Se]$_{1.2}$ | [Li$_{0.15}$Fe$_{0.62}$Al$_{0.23}$(OH)$_2$][Fe$_{0.97}$Se]$_{1.2}$ |
| formula wt | 230.24 | 235.64 |
| Temp/K | 298 | 298 |
| cryst syst | triclinic | triclinic |
| space group | C 1 | C 1 |
| $a$/Å | 5.5414(3) | 5.5300(7)) |
| $b$/Å | 15.393(7) | 15.569(2) |
| $c$/Å | 11.0879(4) | 11.3652(9) |
| $\alpha$/deg | 90.0072(14) | 89.9971(7) |
| $\beta$/deg | 93.727(9) | 94.582(13) |
| $\gamma$/deg | 89.9950(9) | 90.0004(5) |
| $V$/Å$^3$ | 943.8(4) | 975.4(2) |
| Z | 2 | 2 |
| $\rho_{calcd}$/g cm$^{-3}$ | 4.017 | 3.977 |
| F(000) | 1041.9 | 1065.9 |
| Radiation | Cu K$\alpha$, $\lambda$ = 1.54184 Å | Cu K$\alpha$, $\lambda$ = 1.54184 Å |
| 2$\theta$ range/deg | 5-120 | 5-120 |
| h,k,l max | 6, 7, 12 | 6, 7, 12 |
| no. of rflns collected | 2804 | 2916 |
| $R_{wp}$ | 0.1100 | 0.1215 |
| R$_p$ | 0.0810 | 0.0894 |
| $\chi^2$ | 5.486 | 6.180 |

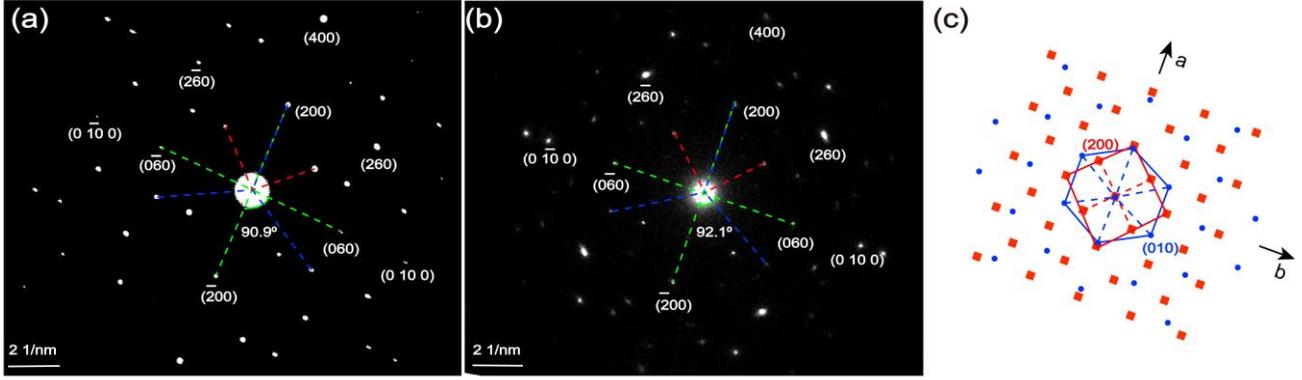

**Figure 2.** (a) Electron diffraction pattern along the [001] zone axis of the as-synthesized [(Fe,Al)(OH)$_2$][FeSe]$_{1.2}$. (b) Electron diffraction pattern along the [001] zone axis of the lithiated [(Fe,Al)(OH)$_2$][FeSe]$_{1.2}$. (c) A sketch of electron diffraction pattern of [(Fe,Al)(OH)$_2$][FeSe]$_{1.2}$, red squares represent reflections from tetragonal sublattice, blue circles represent reflections from hexagonal sublattice. The reflections in (a)(b) represent the reflections of the [(Fe,Al)(OH)$_2$][FeSe]$_{1.2}$. The reflections in (c) represent the reflections of the sublattices.

To give more information of the *ab* plane about the structure of the layered heterostructure [(Fe,Al)(OH)$_2$][FeSe]$_{1.2}$, we have also performed electron diffraction (ED) pattern along the [001] zone axis. Figure 2a and b show the ED patterns of the as-synthesized sample and the lithiated sample with the (*hk*0) reflections which are difficult to resolve from powder XRD, respectively. The angle between the cross-sections connecting the (200) and ($\bar{2}$00) and (060) and (0$\bar{6}$0) reflections from Figure 2a is about 90.9°, and the average lattice parameters *a* and *b* obtained from Figure 2a are 5.59(5) Å and 15.65(1) Å, which are in accordance with the XRD refinement of the as-synthesized sample, respectively. The angle between the cross-sections connecting the (200) and ($\bar{2}$00) and (060) and (0$\bar{6}$0) reflections from Figure 2b is about 92.1°, and the average lattice parameters *a* and *b* obtained from Figure 2b are 5.33(7) Å and 15.86(4) Å, which are also in accordance with the XRD refinement of the lithiated sample, respectively. The reflections in the patterns shown in Figure 2a and b can be fit into two sublattices of one tetragonal and the other hexagonal, which are very similar to the ED patterns of the natural tochilinite.[18] Figure 2c shows a sketch of ED pattern, in which the red squares represent reflections from the FeSe sublattice with tetragonal symmetry, and the blue circles represent reflections from the (Fe,Al)(OH)$_2$ layer with hexagonal symmetry, being similar to the sketch of ED pattern for the natural tochilinite reported by Organova.[18] From the sketch of ED pattern, we can see the relationship between the two sublattices. Along the a-axis direction of the [(Fe,Al)(OH)$_2$][FeSe]$_{1.2}$ heterostructure, the relationship $\sqrt{2}a_{FeSe} = \sqrt{3}\, a_{(Fe,Al)(OH)_2}$ is satisfied. $\sqrt{2}$ times the *a* axis of FeSe is approximately equal to $\sqrt{3}$ times the *a* axis of (Fe,Al)(OH)$_2$. Along the b-axis direction of the [(Fe,Al)(OH)$_2$][FeSe]$_{1.2}$ heterostructure, the relationship $3a_{sq} = 5a_{(Fe,Al)(OH)_2}$ is fulfilled. 3 times the diagonal of a unit FeSe cell is approximately equal to 5 times the *a* axis of (Fe,Al)(OH)$_2$. So a superlattice is formed in the ab plane of the [(Fe,Al)(OH)$_2$][FeSe]$_{1.2}$ heterostructure. The detailed discussion of the superlattices is shown in the supporting information. From the a-axis and b-axis relationships of the [(Fe,Al)(OH)$_2$][FeSe]$_x$ heterostructure, we can get the ideal formula of the heterostructure is [(Fe,Al)(OH)$_2$][FeSe]$_{1.2}$, like the natural tochilinite.[16] This is the reason why the layered [(Fe,Al)(OH)$_2$][FeSe]$_{1.2}$ heterostructure can be synthesized successfully. Because there exists a mismatch between the FeSe layer and (Fe,Al)(OH)$_2$ layer in [(Fe,Al)(OH)$_2$][FeSe]$_{1.2}$, the red square spot and the blue circle spot are not completely overlapped in the (200) reflection of the [(Fe,Al)(OH)$_2$][FeSe]$_{1.2}$ as shown in Figure 2c. The results of the ED patterns and the XRD refinements before and after lithiation indicate that the newly synthesized [(Fe,Al)(OH)$_2$][FeSe]$_{1.2}$ is a layered heterostructure material, which is formed by alternating stacking of tetragonal anti-PbO-type FeSe layer and hexagonal (Fe,Al)(OH)$_2$ layer.

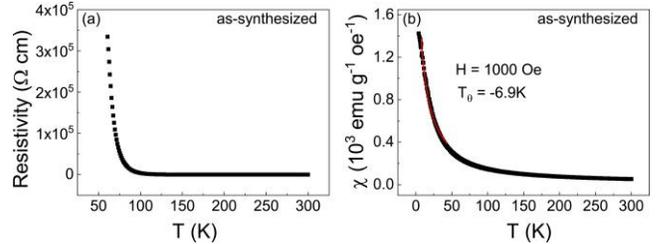

**Figure 3.** (a) The temperature dependent resistivity of the as-synthesized [(Fe,Al)(OH)$_2$][FeSe]$_{1.2}$. (b) The temperature dependent magnetic susceptibility of the as-synthesized [(Fe,Al)(OH)$_2$][FeSe]$_{1.2}$.

The resistivity and magnetic susceptibility of the as-synthesized [(Fe,Al)(OH)$_2$][FeSe]$_{1.2}$ sample are shown in Figure 3. The resistivity of the as-synthesized sample in Figure 3a shows an insulator-like behavior. Such insulating behavior is similar to that of $K_2Fe_4Se_5$ and solid-state synthesized (Li,Fe)OHFeSe, which is attributed to the large amount of Fe vacancies in the FeSe layer.[24,27] Figure 3b shows the temperature dependent magnetic susceptibility of the as-synthesized sample. As shown in Figure 3b, a Curie-Weiss paramagnetism-like behavior is observed. It suggests the existence of local moment due to the Fe vacancies in the FeSe layer. Besides, we can get the Weiss temperature of ∼ -6.9 K via fitting the data with the Curie-Weiss law in the temperature from 2 - 40 K. It suggests the antiferromagnetic exchange coupling in the as-synthesized sample.

Figure 4a shows the temperature dependent resistivity of the lithiated sample. It displays a sharp transition around 40 K. The $T_c^{onset}$ is 40 K and the $T_c^{zero}$ is 38 K. A small hump in resistivity is observed in Figure 4a and the similar hump behavior can also observed in $A_xFe_2Se_2$ (A = K and Cs)[28] and $(Li_{1-x}Fe_x)$OHFeSe.[29] As mentioned in the above references, the hump in resistivity should arise from the defects within the conducting FeSe layers. The reason why no hump is found in the resistivity curve of Figure 3a may be that the hump is a weak hump, and the resistivity at 300 K of the sample before lithiation is about ten times that of the sample after lithiation. In addition, comparing Figure 3a with Figure 4a, the normal state resistivity at 300 K of the sample after lithiation is about one-tenth of that of the sample before lithiation. This decrease is consistent with lithiation process for the lithiation process can eliminate the Fe vacancies in the FeSe layer and introduce carrier.[30,31] The temperature dependent magnetic susceptibility of the lithiated sample is shown in Figure 4b. The measurement is performed under an external field $H$ = 10 Oe. The lithiated sample shows a diamagnetic transition around 38 K, which is consistent with the resistivity measurement. The shielding fraction is more than 80%, indicating that the superconductivity is a bulk superconductivity after lithiation. The emergence of superconductivity in the lithiated sample also proves that the lithiation can effectively eliminate the Fe vacancies in the FeSe layer, which is quite similar the lithiation process in $Li_{1-x}Fe_x(OH)Fe_{1-y}Se$.[12,13] The $T_c$ is nearly the same as that of (Li,Fe)OHFeSe even though the structure of [(Fe,Al)(OH)$_2$][FeSe]$_{1.2}$ is quite different from that of (Li,Fe)OHFeSe. It further indicates that the $T_c$ of the FeSe based superconductors is mainly affected by doping effect of the FeSe plane.[30,31] The resistivity measurements at various magnetic fields are shown in Figure 4c. The $T_c$ decreases with increasing magnetic field as shown in Figure 4c. The temperature dependence of the upper critical field $H_{c2}$ (T) for the lithiated sample is shown in Figure 4d. The upper critical field at $T$ = 0 K can be determined by the Werthamer-Helfand-Hohenberg (WHH) equation[32] $H_{c2}(0) = 0.693[-(dH_{c2}/dT)]_{T_c}T_c$. From the Figure 4d, we can have $[-(dH_{c2}/dT)]_{T_c}$ = 2.18 T/K and $T_c$ = 40 K. Then the $H_{c2}(0)$ can be estimated to be 60.4 T, which is smaller than that of (Li,Fe)OHFeSe single crystal.[11,33]

There exist some common features in the structures of [(Fe,Al)(OH)$_2$][FeSe]$_{1.2}$ and (Li,Fe)OHFeSe. Both of them contain the FeSe layer and the hydroxide layer. Fe vacancies are to appear in the FeSe layer in the above two materials, and the superconductivity appears after the Fe vacancies are filled. The superconducting transition temperature $T_c$ of the two superconductors is much higher than 8 K of the pristine FeSe due to the charge transfer from the hydroxide layer to the FeSe layer. The difference in the structure of these two materials is that both of the LiOH and FeSe are tetragonal structures in (Li,Fe)OHFeSe, while the (Fe,Al)(OH)$_2$ has a hexagonal structure in [(Fe,Al)(OH)$_2$][FeSe]$_{1.2}$. There exists a mismatch between the FeSe and (Fe,Al)(OH)$_2$. The lattices of the FeSe and (Fe,Al)(OH)$_2$ is quasi-commensurate like the case of natural tochilinite. This is the reason why [(Fe,Al)(OH)$_2$][FeSe]$_{1.2}$ single crystal is more difficult to obtain relative to (Li,Fe)OHFeSe single crystal. In nature, the sizes of natural tochilinite needles or fibers are about 10 mm in length and 10-30 μm in diameter,[20] and the tochilinite single crystal cannot be obtained in laboratory so far. Although there exists a mismatch between the tetragonal FeSe layer and the hexagonal (Fe,Al)(OH)$_2$ layer to form layered heterostructure [(Fe,Al)(OH)$_2$][FeSe]$_{1.2}$, the $T_c$ for both of the layered [(Fe,Al)(OH)$_2$][FeSe]$_{1.2}$ heterostructure and the (Li,Fe)OHFeSe single crystal are about 40 K. These results indicate that the superconductivity of FeSe is very robust. The $T_c$ of the FeSe based superconductors is mainly affected by doping effect to the FeSe plane.[30,31]

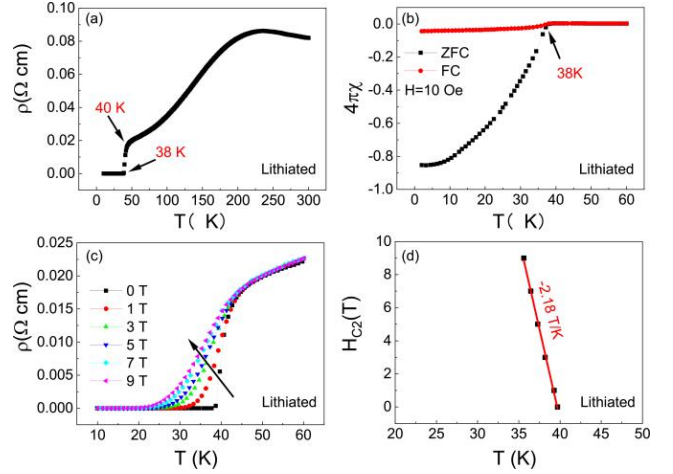

**Figure 4.** (a) The temperature dependent resistivity of the lithiated [(Fe,Al)(OH)$_2$][FeSe]$_{1.2}$. (b) The temperature dependent magnetic susceptibility of the lithiated [(Fe,Al)(OH)$_2$][FeSe]$_{1.2}$. (c) The resistivity measurements in magnetic fields (H) of 0, 1, 3, 5, 7 and 9 T below 60 K for the lithiated [(Fe,Al)(OH)$_2$][FeSe]$_{1.2}$. (d) The temperature dependence of the upper critical field $H_{c2}$(T) for the lithiated [(Fe,Al)(OH)$_2$][FeSe]$_{1.2}$.

## CONCLUSIONS

In summary, we successfully synthesize a new layered superconductor of [(Fe,Al)(OH)$_2$][FeSe]$_{1.2}$ using a hydrothermal ion-exchange technique. The XRD and ED results of the [(Fe,Al)(OH)$_2$][FeSe]$_{1.2}$ show that this layered material is composed of the alternating stacking of tetragonal anti-PbO-type FeSe layer and hexagonal (Fe,Al)(OH)$_2$ layer. The as-synthesized [(Fe,Al)(OH)$_2$][FeSe]$_{1.2}$ is non-superconducting due to the Fe vacancies in the FeSe layer. The superconductivity with $T_c$=40 K is achieved after lithiation to eliminate the Fe vacancies in the FeSe layer. This new layered heterostructure is conducive to the exploration of new iron-base superconductors.

*chenxh@ustc.edu.cn

## ACKNOWLEDGMENT

This work was supported by the National Natural Science Foundation of China (11888101 and 11534010), the National Key Research and Development Program of the Ministry of Science and Technology of China ( 2017YFA0303001, 2016YFA0300201 and 2019YFA0704901), the Strategic Priority Research Program of Chinese Academy of Sciences (XDB25000000), Anhui Initiative in Quantum Information Technologies (AHY160000), the Science Challenge Project of China (Grant No. TZ2016004), and the Key Research Program of Frontier Sciences, CAS, China (QYZDYSSW-SLH021).

## REFERENCES

(1) Kamihara, Y.; Watanabe, T.; Hirano, M.; Hosono, H. Iron-based layered superconductor LaO$_{1-x}$F$_x$FeAs (x = 0.05-0.12) with $T_c$ = 26 K. *J. Am. Chem. Soc.* **2008**, *130*, 3296-3297.
(2) Chen, X. H.; Wu, T.; Wu, G.; Liu, R. H.; Chen, H.; Fang, D. F. Superconductivity at 43 K in SmFeAsO$_{1-x}$F$_x$. *Nature* **2008**, *453*, (7196), 761-762.
(3) Peng, X. L.; Li, Y.; Wu, X. X.; Deng, H. B.; Shi, X.; Fan, W. H.; Li, M.; Huang, Y. B.; Qian, T.; Richard, P.; Hu, J. P.; Pan, S. H.; Mao, H. Q.; Sun, Y. J.; Ding, H. Observation of topological transition in high-$T_c$ superconducting monolayer FeTe$_{1-x}$Se$_x$ films on SrTiO$_3$(001). *Phys. Rev. B* **2019**, *100*, 155134.
(4) Sun, R.; Jin, S.; Gu, L.; Zhang, Q.; Huang, Q.; Ying, T.; Peng, Y.; Deng, J.; Yin, Z.; Chen, X. Intercalating Anions between Terminated Anion Layers: Unusual Ionic S–Se Bonds and Hole-Doping Induced Superconductivity in S$_{0.24}$(NH$_3$)$_{0.26}$Fe$_2$Se$_2$. *J. Am. Chem. Soc.* **2019**, *141*, 13849-13857.
(5) Zhou, X.; Eckberg, C.; Wilfong, B.; Liou, S. C.; Vivanco, H. K.; Paglione, J.; Rodriguez, E. E. Superconductivity and magnetism in iron sulfides intercalated by metal hydroxides. *Chem. Sci.* **2017**, 8, 3781-3788.
(6) McQueen, T. M.; Huang, Q.; Ksenofontov, V.; Felser, C.; Xu, Q.; Zandbergen, H.; Hor, Y. S.; Allred, J.; Williams, A. J.; Qu, D.; Checkelsky, J.; Ong, N. P.; Cava, R. J. Extreme sensitivity of superconductivity to stoichiometry in Fe$_{1+\delta}$Se. *Phys. Rev. B* **2009**, *79*, 014522.
(7) Ying, T. P.; Chen, X. L.; Wang, G.; Jin, S. F.; Zhou, T. T.; Lai, X. F.; Zhang, H.; Wang, W. Y. Observation of superconductivity at 30∼46K in A$_x$Fe$_2$Se$_2$ (A= Li, Na, Ba, Sr, Ca, Yb and Eu). *Sci. Rep.* **2012**, *2*, 1-7.
(8) Krzton-Maziopa, A.; Pomjakushina, E. V.; Pomjakushin, V. Y.; Von Rohr, F.; Schilling, A.; Conder, K. Synthesis of a new alkali metal–organic solvent intercalated iron selenide superconductor with $T_c$≈ 45 K. *J. Phys-condens. Mat.* **2012**, *24*, 382202.
(9) Lu, X. F.; Wang, N. Z.; Zhang, G. H.; Luo, X. G.; Ma, Z. M.; Lei, B.; Huang, F. Q.; Chen, X. H. Superconductivity in LiFeO$_2$Fe$_2$Se$_2$ with anti-PbO-type spacer layers. *Phys. Rev. B* **2014**, *89*, 020507.
(10) Lu, X. F.; Wang, N. Z.; Wu, H.; Wu, Y. P.; Zhao, D.; Zeng, X. Z.; Luo, X. G.; Wu, T.; Bao, W.; Zhang, G. H.; Huang, F. Q.; Huang, Q. Z.; Chen, X. H. Coexistence of superconductivity and antiferromagnetism in (Li$_{0.8}$Fe$_{0.2}$)OHFeSe. *Nat. Mater.* **2015**, *14*, 325-329.
(11) Dong, X.; Jin, K.; Yuan, D.; Zhou, H.; Yuan, J.; Huang, Y.; Hua, W.; Sun, J.; Zheng, P.; Hu, W.; Mao, Y.; Ma, M.; Zhang, G.; Zhou, F.; Zhao, Z. (Li$_{0.84}$Fe$_{0.16}$)OHFe$_{0.98}$Se superconductor: Ion-exchange synthesis of large single-crystal and highly two-dimensional electron properties. *Phys. Rev. B* **2015**, *92*, 064515.
(12) Woodruff, D. N.; Schild, F.; Topping, C. V.; Cassidy, S. J.; Blandy, J. N.; Blundell, S. J.; Thompson, A. L.; Clarke, S. J. The parent Li(OH)FeSe phase of lithium iron hydroxide selenide superconductors. *Inorg. Chem.* **2016**, *55*, 9886-9891.
(13) Sun, H.; Woodruff, D. N.; Cassidy, S. J.; Allcroft, G. M.; Sedlmaier, S. J.; Thompson, A. L.; Bingham, P. A.; Forder, S. D.; Cartenet, S.; Mary, N.; Ramos, S.; Foronda, F. R.; Williams, B. H.; Li, X.; Blundell, S. J.; Clarke, S. J. Soft chemical control of superconductivity in lithium iron selenide hydroxides Li$_{1-x}$Fe$_x$(OH)Fe$_{1-y}$Se. *Inorg. Chem.* **2015**, *54*, 1958-1964.
(14) Wiegers, A. G. Misfit layer compounds: structures and physical properties. *Prog. Solid State Ch.* **1966**, *24*, 1-139.
(15) Monceau, P.; Chen, J.; Laborde, Q.; Briggs, A. Anisotropy of the superconducting properties of misfit layer compounds (MX)$_n$(NbX$_2$)$_m$. *Physica B* **1994**, *194*, 2361-2362.
(16) Organova, N. I.; Drits, V. A.; Dmitrik, A. L. Structural study of tochilinite. Part I. The isometric variety. *Sov. Phys. Crystallogr.* **1972**, *17*, 667.
(17) Moroz, L. V.; Kozerenko, S. V.; Fadeev, V. V. The reflectance spectrum of synthetic tochilinite. *Lunar Planet. Sci. XXVIII.* Lunar Planet. Inst., Houston. #1288 (abstr.) **1997**, *983*.
(18) Organova, N. I.; Gorshkov, A. I.; Dikov, Y. P.; Kul'bachinskiy, V. A.; Laputina, I. P.; Sivtsov, A. V.; Sluzhenikin, S. F.; Ponomarenko, A. I. New data on tochilinite. *Int. Geol. Rev.* **1988**, *30*, 691-705.
(19) Peng, Y.; Xu, L.; Xi, G.; Zhong, C.; Lu, J.; Meng, Z.; Li, G.; Zhang, S.; Zhang, G.; Qian, Y. An experimental study on the hydrothermal preparation of tochilinite nanotubes and tochilinite–serpentine-intergrowth nanotubes from metal particles. *Geochim. Cosmochim. Ac.* **2007**, *71*, 2858-2875.
(20) Kakos, G. A.; Turney, T. W.; Williams, T. B. Synthesis and structure of tochilinite: a layered metal hydroxide/sulfide composite. *J. Solid State Chem.* **1994**, *108*, 102-111.
(21) Larson, A. C.; Von Dreele, R. B. General Structure Analysis System (GSAS). Los Alamos National Laboratory Report LAUR 86− 748; The University of California: 1994.
(22) Toby, B. H. EXPGUI, a graphical user interface for GSAS. *J. Appl. Crystallogr.* **2001**, *34*, 210−213.
(23) Pekov, I. V.; Sereda, E. V.; Polekhovsky, Y. S.; Britvin, S.; N. Chukanov, N. V.; Yapaskurt, V. O.; Bryzgalov, I. A. Ferrotochilinite, 6FeS· 5Fe(OH)$_2$, a new mineral from the Oktyabr'sky deposit, Noril'sk district, Siberia, Russia. *Geol. Ore Deposit+.* **2013**, *55*, 567-574.
(24) Hu, G. B.; Wang, N. Z.; Shi, M. Z.; Meng, F. B.; Shang, C.; Ma, L. K.; Luo, X. G.; Chen, X. H. Superconductivity in solid-state synthesized (Li, Fe)OHFeSe by tuning Fe vacancies in FeSe layer. *Phys. Rev. Mater.* **2019**, *3*, 064802.
(25) Shannon, R. D. Revised Effective Ionic Radii and Systematic Studies of Interatomic Distances in Halides and Chalcogenides. *Acta Crystallogr. Sect. A* **1976**, *32*,751-767.
(26) Ksenofontov, V.; Wortmann, G.; Medvedev, S. A.; Tsurkan, V.; Deisenhofer, J.; Loidl, A.; Felser, C. Phase separation in superconducting and antiferromagnetic Rb$_{0.8}$Fe$_{1.6}$Se$_2$ probed by Mössbauer spectroscopy. *Phys. Rev. B* **2011**, *84*, 180508.
(27) Song, Y. J.; Wang, Z.; Wang, Z. W.; Shi, H. L.; Chen, Z.; Tian, H. F.; Chen, G. F.; Yang, H. X.; Li, J. Q. Phase transition, superstructure and physical properties of K$_2$Fe$_4$Se$_5$. *Europhys. Lett.* **2011**, *95*, 37007.
(28) Ying, J. J.; Wang, X. F.; Luo, X. G.; Li, Z. Y.; Yan, Y. J.; Zhang, M.; Wang, A. F.; Cheng, P.; Ye, G. J.; Xiang, Z. J.; Liu, R. H.; Chen, X. H. Pressure effect on superconductivity of A$_x$Fe$_2$Se$_2$ (A= K and Cs). *New J. Phys.* **2011**,*13*, 033008.
(29) Ryu, G. H.; Zhang, G. Y.; Yu, G.; Chou, M. M. C.; Lin, C. T. Anisotropic Behaviors of (Li$_{1-x}$Fe$_x$)OHFeSe Superconducting Single Crystals. *IEEE T. Appl. Supercon.* **2018**, 28, 1-5.
(30) Lei, B.; Cui, J. H.; Xiang, Z. J.; Shang, C.; Wang, N. Z.; Ye, G. J.; Luo, X. G.; Wu, T.; Sun, Z.; Chen, X. H. Evolution of high-temperature superconductivity from a Low-$T_c$ phase tuned by carrier concentration in FeSe thin flakes. *Phys. Rev. Lett.* **2016**, *116,* 077002.
(31) Cui, Y.; Zhang, G.; Li, H.; Lin, H.; Zhu, X.; Wen, H. H.; Wang, G.; Sun, J. Ma, M.; Li, Y.; Gong, D.; Xie, T.; Gu, Y.; Li, S.; Luo, H.; Yu, P.; Yu, W. Protonation induced high-$T_c$ phases in iron-based superconductors evidenced by NMR and magnetization measurements. *Sci. Bull.* **2018**, *63*, 11-16.
(32) Werthamer, N. R.; Helfand, E.; Hohenberg, P. C. Temperature and purity dependence of the superconducting critical field, $H_{c2}$. III. Electron spin and spin-orbit effects. *Phys. Rev.* **1966**, *147*, 295.

(33) Huang, Y.; Feng, Z.; Ni, S.; Li, J.; Hu, W.; Liu, S.; Mao, Y.; Zhou, H.; Zhou, F.; Jin, K.; Wang, H.; Yuan, J.; Dong, X.; Zhao, Z. Superconducting (Li,Fe)OHFeSe film of high quality and high critical parameters. *Chinese Phys. Lett.* **2017**, *34*, 077404.